\documentclass{kluwer}
\usepackage{epsfig}
\usepackage{epsf}
\def\ltsima{$\; \buildrel < \over \sim \;$}
\def\simlt{\lower.5ex\hbox{\ltsima}}\begin{document}
\begin{article}
\begin{opening}
\title{The Accretion-Ejection Instability in X-ray Binaries}            
\author{P. \surname{Varni\`ere}\footnote{\email{varniere@cea.fr}}} \author{M.
\surname{Tagger}} \institute{Service d'Astrophysique (CNRS URA 2052)
,CEA/Saclay 91191 Gif sur Yvette}
\begin{abstract} 
The Accretion-Ejection Instability (AEI), which can occur in magnetized
disks near equipartition, is a good candidate to explain the
low-frequency QPO in black-hole binaries.  Here we present analytical
work concerning the behavior of QPO frequency and the emission of Alfv\'en
waves from the disk to the corona.                                              
\end{abstract}
\keywords{Accretion, accretion disks - Instabilities - 
MHD Waves - Galaxies: jets}

\end{opening}

\section{Introduction}

        The AEI is a spiral instability, similar to galatic spirals but
        driven by magnetic stress rather than self-gravity.  It occurs
        in the inner region of an accretion disk threaded by a vertical
        magnetic field of the order of equipartition with the gas
        pressure.  The spiral extracts energy and angular momentum from
        the disk, causing accretion, and stores them in a Rossby vortex
        at its corotation radius (see also the contributions of Tagger
        and Caunt, these proceedings).  This vortex then leaks energy
        and angular momentum as Alfv\'en waves to the corona, where it can
        power a wind or a jet. For more details see Tagger,M.\& Pellat,R.
	 (1999).

\section{QPO Frequency in Pseudo-Newtonian Potential}

        Relativistic effects change the rotation curve of the disk near
        the last stable orbit.  They allow the existence of an Inner
        Lindblad Resonance (ILR) for the $m=1$ mode.  This changes the
        properties of the 1-armed spiral (best candidate to explain the
        QPO), and therefore  its frequency.
 
        We have studied, using a pseudo-newtonian potential, the
        properties of the instability when the disk inner radius
        $r_{int}$ approaches the Last Stable Orbit at $r_{LSO}$.  When
        $r_{int}$ is large the QPO frequency varies as $\omega \propto
        r^{-3/2}$; but as $r_{int}$ approaches $r_{LSO}$ the correlation
        changes and becomes {\em positive} (the QPO frequency decreasing
        with a decreasing radius) when $r_{int} < 1.4 r_{LSO}$, {\em
        i.e.} when the $m=1$ mode has an Inner Lindblad Resonance in the
        disk \cite{VAR}.
	We consider this as a possible explanation for the
        positive correlation found (contrary to other sources) in GRO
	J$1655$ as shown in these proceedings and Rodriguez,J., 
	Varni\`ere,P.\& Tagger,M. (2000).
\begin{figure}
\begin{tabular}{cc}
\hspace*{0.75cm}\epsfig{file=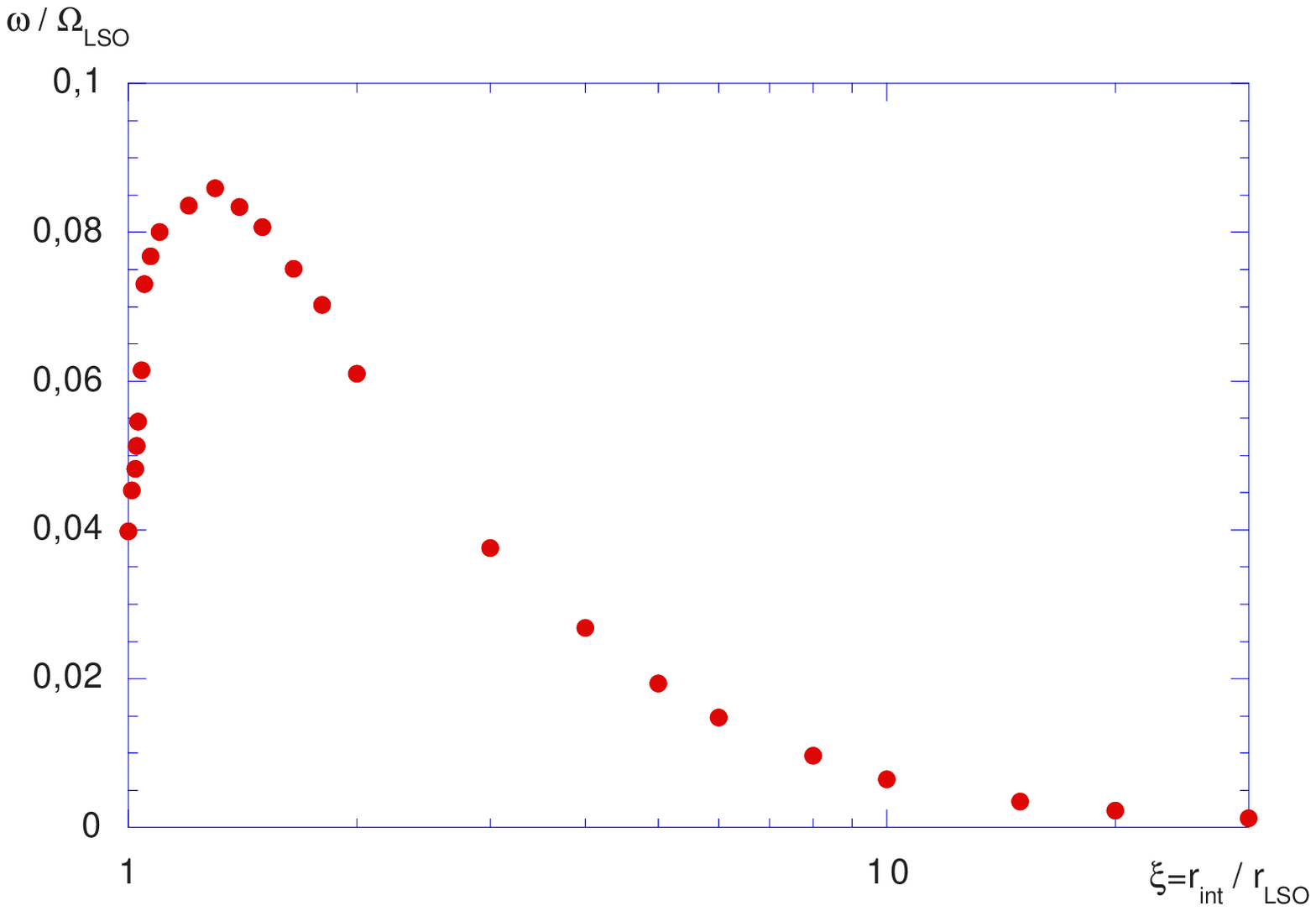,width=4.5cm}&
\hspace*{0.75cm}\epsfig{file=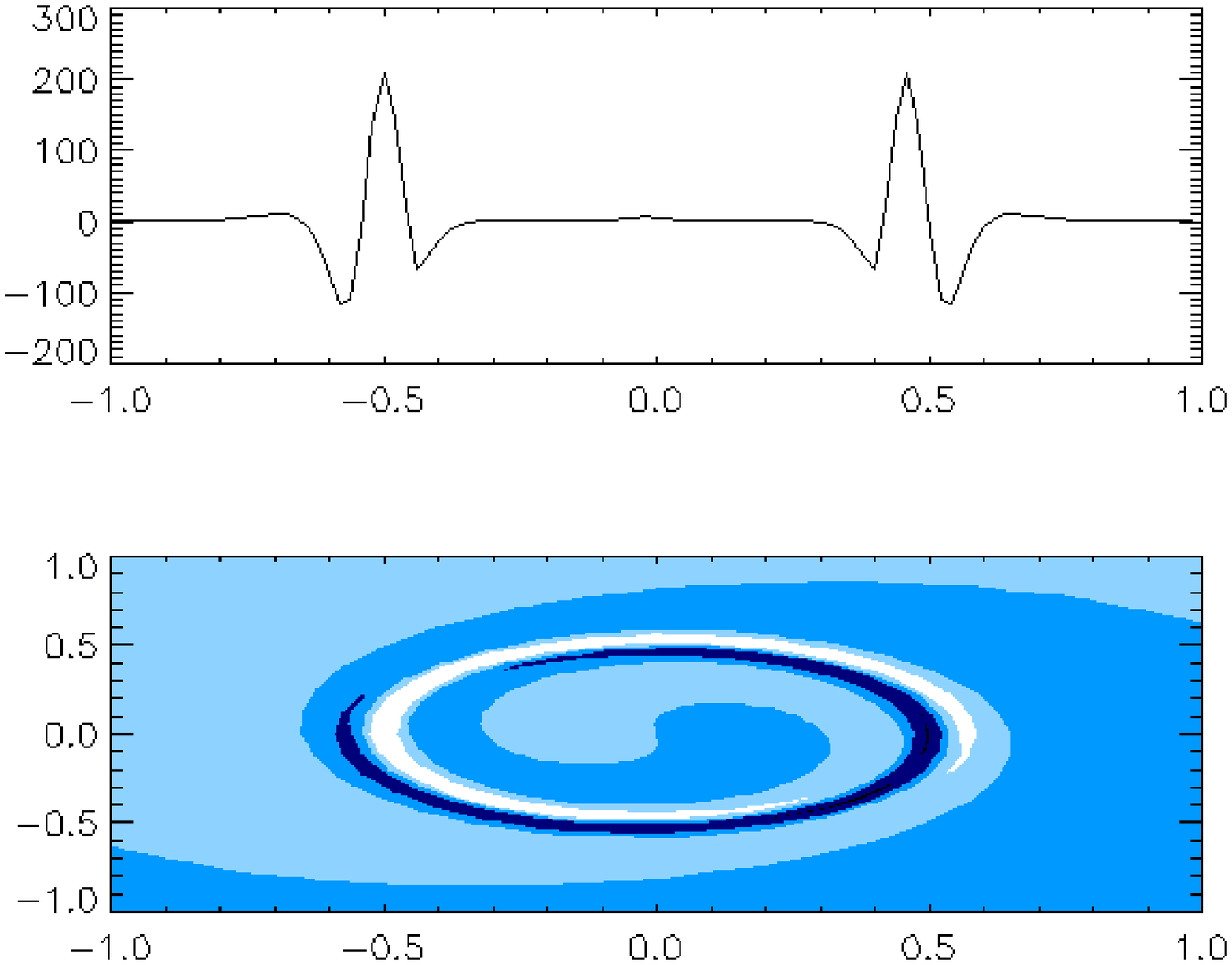,width=4cm}
\end{tabular}
\caption{(left) the QPO frequency as a function of the disk inner radius
$r_{int}$.  The frequency-radius correlation is reversed when   
 $r_{int}\simlt 1.4 r_{LSO}$.  (right) radial profile and spatial structure of the flux of
Alfv\'en waves emitted to the corona.}
\end{figure}
\section{Emission of Alfv\'en Waves}

The Rossby vortex twists the footpoint of the field lines threading the
disk.  If the disk has a low density corona this twisting will be
propagated upward as Alfv\'en waves.  The energy and angular momentum
extracted from the disk will thus be transfered to the corona where they
can power a wind or jet.  We study this with a variational form   
              
\begin{eqnarray*}
F=[\mbox{energy of the waves}]
+i &[\mbox{outgoing spiral}
+\mbox{ vortex}\nonumber \\
  &+k_z\  \mbox{Alfv\'en Waves}]\nonumber
\end{eqnarray*}
where imaginary terms correspond to amplification or damping of the
instability.  From the numerical solution we compute the flux of the
emitted Alfv\'en waves.  The result, plotted in Figure 1, shows that it
peaks at the corotation radius, where the Rossby vortex is localized.
\section{Conclusions}
The frequency of the AEI, and its variations with the disk inner radius,
make it a very good candidate to explain the ``ubiquitous'' QPO. The
accretion energy and momentum extracted from the disk are propagated to
the corona as Alfv\'en waves.

\end{article}
\end{document}